\documentclass[12pt]{iopart}

\usepackage{graphicx}

\newcommand{\ket}[1]{\left| #1 \right>} 
\newcommand{\bra}[1]{\left< #1 \right|} 

\begin{document}

\title[$GW$ quasiparticle band gaps of anatase TiO$_2$ starting from DFT+$U$]
{$\mathbf{GW}$ quasiparticle band gaps of anatase TiO$_\mathbf{2}$ starting from DFT+$\mathbf{U}$}

\author{Christopher E. Patrick and Feliciano Giustino}

\address{Department of Materials, University of Oxford, Parks Road,
Oxford OX1 3PH, United Kingdom}
\ead{feliciano.giustino@materials.ox.ac.uk}
\begin{abstract}
We investigate the quasiparticle band structure of anatase TiO$_2$, a wide gap
semiconductor widely employed in photovoltaics and photocatalysis. We obtain
$GW$ quasiparticle energies starting from density-functional theory (DFT) calculations
including Hubbard $U$ corrections. Using a simple iterative procedure we determine
the value of the Hubbard parameter yielding a vanishing quasiparticle
correction to the fundamental band gap of anatase TiO$_2$. The band gap (3.3~eV) calculated
using this optimal Hubbard parameter is smaller than the value obtained by
applying many-body perturbation theory to standard DFT eigenstates and eigenvalues
(3.7~eV). We extend our analysis to the rutile polymorph of TiO$_2$ and reach
similar conclusions. Our work highlights the role of the starting
non-interacting Hamiltonian in the calculation of $GW$ quasiparticle energies
in TiO$_2$, and suggests an optimal Hubbard parameter for future calculations.
\end{abstract}

\maketitle

The wide gap semiconductor TiO$_2$ is the subject 
of a vast research effort owing to its importance to
many areas of technology, ranging from photovoltaics and photocatalysis to
sensing devices, protective coatings and batteries \cite{Diebold2003}.
As an example, nanostructured 
anatase TiO$_2$ films sensitized with molecular dyes are the fundamental 
building blocks of dye-sensitized solar cells \cite{ORegan1991}.
In order to understand and optimize the photovoltaic and photocatalytic 
properties of TiO$_2$ it is important to complement experimental spectroscopies 
with reliable calculations of electronic excitations in this material
and its interfaces.

Band structures calculated within standard DFT underestimate the fundamental
band gap of anatase TiO$_2$ by about 1~eV. The underestimation of the band gap 
is accompanied by errors of similar magnitude in the individual addition and removal 
energies, and raises concerns about the calculation of energy-level alignments 
and charge injection dynamics at interfaces involving TiO$_2$ \cite{Martsinovich2011}.

In this context accurate quasiparticle methods for studying electronic
excitations, such as the $GW$ approximation in its various implementations 
\cite{Hedin1965,Hybertsen1986,Onida2002,Rinke2005,Schilfgaarde2006}, still face two
limitations. (i) On the one hand these methods are computationally demanding, and cannot 
be used yet for complex systems such as defects and interfaces. 
For example the minimal atomistic models required to describe donor/acceptor 
interfaces for solar cells involve several hundreds of atoms \cite{Patrick2011,
Patrick2011-AFM, Angelis20112}, which are beyond the reach of current $GW$ implementations.
(ii) On the other hand, there is the issue of whether $GW$ calculations
should be performed at the $G_0W_0$ level, or include some form of self-consistency 
\cite{Schilfgaarde2006,Bruneval2006} or off-diagonal matrix elements of
the self-energy \cite{Bruneval2006}.

In this work we attempt to address the latter point by combining DFT+$U$ and $G_0W_0$
calculations, in the spirit of
Kioupakis \emph{et al} \cite{Kioupakis2008}.
The rationale for this approach is that the use of DFT+$U$ as the noninteracting 
Hamiltonian should provide a better starting point for perturbation theory, thereby 
legitimizing the use of the $G_0W_0$ approximation.
The Hubbard parameter is determined in such a way
that a $G_0W_0$ calculation starting from the DFT+$U$ band structure yields
a vanishing correction to the fundamental band gap.
This is achieved through an iterative procedure consisting of
a sequence of $G_0W_0$ and DFT+$U$ calculations, and  can
be seen as a simple and inexpensive way to mimic self-consistent $GW$.
Using this procedure we obtain a band gap of 3.3 eV for anatase TiO$_2$. 
When comparing the band structures obtained from DFT+$U$ with or without quasiparticle
corrections, we find that they are in reasonable agreement only near the band edges, 
and that the individual shifts of the bands determined from DFT+$U$ are always
larger than the corresponding quasiparticle shifts.
Interestingly we find that the optimal value of $U$ thus determined 
is transferable to the rutile polymorph of TiO$_2$.

We perform DFT calculations using the generalized gradient
approximation of
Perdew \emph{et al} \cite{Perdew1996}
and norm-conserving
pseudopotentials \cite{Fuchs1999}. We explicitly include the 3$s$ and 3$p$ semicore
states of Ti in our calculations. The electronic wavefunctions are
expanded in a plane waves basis set with a kinetic energy cutoff of 200~Ry,
and the Brillouin zone of TiO$_2$ is sampled using a 6$\times$6$\times$6 
Monkhorst-Pack mesh. All DFT calculations are performed using the 
\texttt{Quantum ESPRESSO} software package \cite{quantumespresso}.
For definiteness we set the lattice parameters and atomic positions 
within the unit cell to the experimental values 
\footnote{Our optimized lattice parameters 
  are $a$=3.83~\r A\ and $c$=9.84~\r A 
  and the internal parameter is $u$=0.206}
 \cite{Horn1972}.
In order to perform DFT+$U$ calculations \cite{Ansimov1991} we adopt 
the simplified rotationally invariant formulation of 
Cococcioni and de Gironcoli \cite{Cococcioni2005},
with the Hubbard-like corrections included using the atomic Ti 3$d$ 
pseudo-wavefunctions for the projectors. It is worth stressing that,
as in every DFT+$U$ calculation, the value of $U$ determined in this work 
is linked to the particular choice of the projectors \cite{Pickett1998}.
Quasiparticle $GW$ calculations are performed using the scheme of 
Hybertsen and Louie \cite{Hybertsen1986}.
The frequency dependence of the screened Coulomb interaction is described
using the Godby-Needs plasmon-pole model \cite{Godby1989}, by evaluating
the static polarizability as well as the polarizability at the imaginary frequency
of 23~eV. 
  \begin{figure}
  \includegraphics{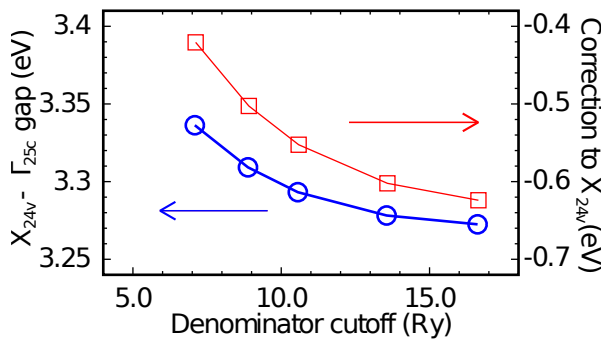}
  \caption{
  Convergence of $GWU$ calculations with the energy denominator cutoff in the
  sum-over-states expansions of the polarizability and the Green's function
  of anatase TiO$_2$: $X_{24v}$-$\Gamma_{25c}$ gap (circles, left scale), 
  and quasiparticle correction to the $X_{24v}$ state (squares, right scale). 
  The quasiparticle corrections were obtained by starting from a DFT+$U$ 
  calculation with $U$=7.5~eV. The lines are guides to the eye.
  \label{fig:gap_convergence}
  }
  \end{figure}
For the screened Coulomb interaction we use a uniform and 
unshifted 4$\times$4$\times$4 Brillouin zone mesh.
The exchange and the correlation
components of the self-energy are described using plane waves basis sets
with kinetic energy cutoffs of 80~Ry and 10~Ry, respectively.
In the expansion over conduction states we include 568 unoccupied bands for calculating
both the polarizability and the Green's function.
The energy denominator cutoff, i.e.\ the energy of the topmost band used in this expansion
referred to the valence band maximum, is 16.6~Ry.
Figure~\ref{fig:gap_convergence} shows the convergence of the fundamental gap with
the denominator cutoff. Increasing the denominator cutoff from 10.6~Ry (296 conduction 
bands) to 16.6~Ry (568 bands) reduces the band gap by 0.02~eV.
The individual corrections to the band edges converge more slowly, as shown in Figure~\ref{fig:gap_convergence}
for the highest occupied state at the $X$ point.
All the $GW$ calculations are performed using the \texttt{SaX} code \cite{MartinSamos2009}.

The calculation of $GW$ quasiparticle energies using DFT+$U$ as the starting 
non-interacting Hamiltonian is performed as follows. We calculate the 
$G_0W_0$ self-energy $\hat{\Sigma}$ using eigenstates and eigenvalues 
obtained from a DFT+$U$ calculation. Then, for each Kohn-Sham state $\psi$ 
of energy $\epsilon$, the quasiparticle energy $E$ is obtained from:
  \begin{equation}
  E = \epsilon + Z(\epsilon)
  \bra{\psi}\hat{\Sigma}(\epsilon)
  -V_{\rm xc} - \hat{V}_U\ket{\psi},
  \label{eqn:QPU}
  \end{equation}
where $Z$ is the quasiparticle renormalization,
$V_{\rm xc}$ is the DFT exchange and correlation potential,
and $\hat{V}_U$ is the Hubbard term
in the Kohn-Sham Hamiltonian. The Hubbard contribution 
to the quasiparticle shift is conveniently evaluated as:
  \begin{equation}
  \bra{\psi}\hat{V}_U \ket{\psi}= \epsilon - \bra{\psi} \hat{H}_0 \ket{\psi},
  \end{equation}
where $\hat{H}_0$ is the Kohn-Sham Hamiltonian without the Hubbard term
(but with the DFT+$U$ density).
In the following we will refer to this procedure simply as $GWU$ method.
We have compared the present approach with the method described by
Jiang \emph{et al} \cite{Jiang2010},
where the dynamical contributions are evaluated at the energy 
$\epsilon-\bra{\psi}\hat{V}_U \ket{\psi}$, and found the differences
to be negligible.
In the case of $U$=0 
equation (\ref{eqn:QPU}) yields the standard
$GW$ quasiparticle energies obtained by using perturbation theory on top of DFT.

The use of DFT+$U$ as the starting Hamiltonian for the perturbation
expansion introduces an additional arbitrariness in the procedure, 
relating to the choice of the Hubbard parameter $U$.
Since we are using $GW$ within the framework of perturbation theory, 
the natural choice for the starting Hamiltonian (and hence the value
of $U$) is the one leading to the smallest quasiparticle corrections.
This observation can be formalized into a criterion for choosing the
Hubbard parameter: our optimal $U$ is determined by the condition that,
upon performing a $GWU$ calculation, the fundamental band gap does not change.
This approach has been demonstrated 
for the case of bcc hydrogen \cite{Kioupakis2008}.
The optimal Hubbard parameter is determined by using an iterative
procedure consisting of successive $GWU$ and DFT+$U$ calculations,
until self-consistency is achieved.
  \begin{figure}
  \includegraphics{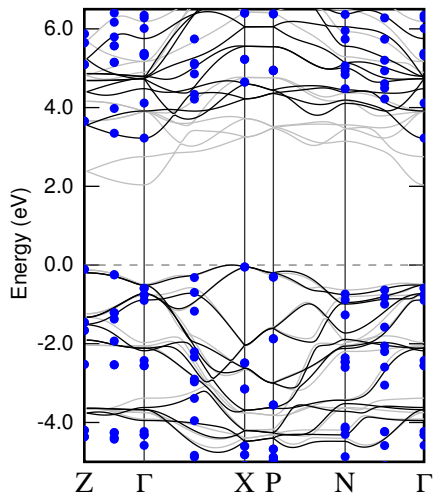}
  \caption{
  Band structure of anatase TiO$_2$ calculated using various methods.
  The gray lines represent standard DFT calculations \cite{Perdew1996},
  the black lines are obtained from DFT+$U$ calculations \cite{Cococcioni2005}
  using our optimal Hubbard parameter $U$=7.5~eV, and the blue disks
  are from $GW$ calculations using DFT+$U$ as the 
  non-interacting Hamiltonian. In each case the top of the valence band 
  is arbitrarily aligned with the zero of the energy axis. 
  \label{fig:U_bandstructure}
  }
  \end{figure}

\begin{figure}
  \includegraphics{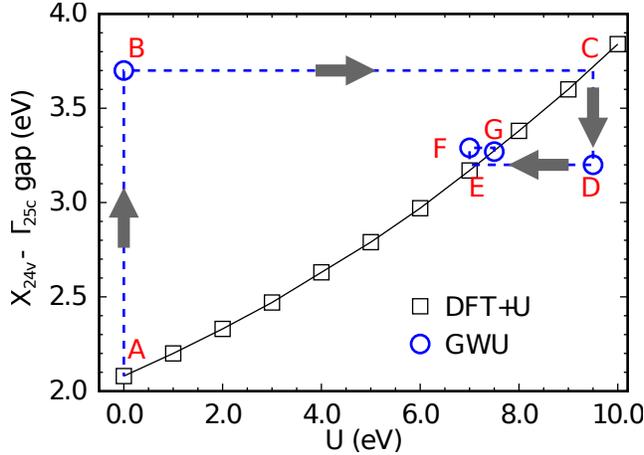}
  \caption{Fundamental band gap ($X_{24v}$-$\Gamma_{25c}$) of anatase TiO$_2$  
  calculated using DFT+$U$ (black squares), and $GWU$ (blue circles), respectively.
  The lines are guides to the eye.
  \label{fig:gap_vs_U}}
  \end{figure}

Figure~\ref{fig:U_bandstructure} shows that 
the band gap of anatase TiO$_2$ is indirect, with the top of the valence band
located near the $X$ point of the Brillouin zone, and the bottom 
of the conduction band at the $\Gamma$ point. 
Since the top of the valence band lies very
close in energy with the states at the $X$ point ($<$60~meV),
for convenience in our analysis we take the states at the $X$ point
as representative of the valence band top. 
We denote the band extrema as $X_{24v}$ and $\Gamma_{25c}$, respectively \cite{Chiodo2010}.
The values of the $X_{24v}$-$\Gamma_{25c}$ gap calculated within DFT+$U$
for a range of values of $U$ between 0 and 10~eV are shown in 
Fig.~\ref{fig:gap_vs_U}, and are in good agreement with previous work \cite{Dompablo2011}.
For $U$=0 we recover the standard DFT band gap of 2.08~eV 
(indicated by $A$ in figure~\ref{fig:gap_vs_U}).

Now we go through the iterative calculation of the band gap.
We calculate the $GW$ quasiparticle corrections starting from the
DFT eigenstates and eigenvalues (i.e.\ from point $A$ in figure~\ref{fig:gap_vs_U}),
and obtain a corrected band gap of 3.70~eV (point $B$ in figure~\ref{fig:gap_vs_U}).
The latter value is in good agreement with previously published works
using the same 
methodology, 3.6--3.8~eV \cite{Chiodo2010, Kang2010, Mowbray2009,Thulin2008}.
We note that the Godby-Needs plasmon-pole model \cite{Godby1989} used in
our calculations produces a band gap in good agreement
(only 0.08~eV larger) with calculations based on the contour 
deformation method \cite{Kang2010}, and does not suffer from the overestimation
of the band gap reported for other plasmon-pole models \cite{Kang2010}.
The success of the Godby-Needs approach 
in reproducing contour deformation results
has recently also been demonstrated for ZnO  \cite{Stankovski2011}.
From figure~\ref{fig:gap_vs_U} we see that the same band gap as in $B$ can be obtained
by using DFT+$U$ and the Hubbard parameter $U$=9.5~eV (point $C$).
By repeating the previous procedure and applying $GW$ corrections to 
the DFT+$U$ eigenstates and eigenvalues obtained with $U$=9.5~eV, we find that
the quasiparticle corrections slightly reduce the band gap to 3.20~eV (point $D$ 
in figure~\ref{fig:gap_vs_U}).
We iterate this procedure until the quasiparticle correction to the band gap
becomes negligible. The iterations generate the converging series 
of Hubbard parameters 0.0$\rightarrow$9.5$\rightarrow$7.0$\rightarrow$7.5~eV, 
and the series of band gaps 
2.08$\rightarrow$3.70$\rightarrow$3.20$\rightarrow$3.29$\rightarrow$3.27~eV. 
By the fourth iteration the DFT+$U$ and $GWU$ band gaps are both 3.27~eV
and the Hubbard parameter is $U$=7.5~eV.
This value of the Hubbard parameter is compatible with the range 
of values 2-8~eV reported in the literature 
for oxides containing Ti, cf.~Nolan \emph{et al} \cite{Nolan2008} and references therein.
It is interesting to check whether the optimal Hubbard parameter $U_{\rm opt}$=7.5 eV
here determined for anatase TiO$_2$ is transferable, i.e.\ (i) whether it generates a  
band structure in agreement with $GWU$, and (ii) whether 
it can also be used for other TiO$_2$ polymorphs.

In order to address the first point we compare in Fig.~\ref{fig:U_bandstructure}
the band structures calculated using DFT+$U$ and those calculated using
the $GWU$ method, both with $U$=$U_{\rm opt}$.
Near the band extrema the two methods yield similar results, although
differences in energy of up to 0.3 eV are observed for several bands.
We also estimate that the $GWU$ calculation yields a valence band width
which is 0.3 eV larger than the starting DFT+$U$ band structure.
These differences are rather small when compared with the broadening of the
valence photoemission spectra reported by Li \emph{et al} \cite{Li20112},
corresponding to $\sim$1~eV.
While the band gaps and band width obtained using DFT+$U$ and $GWU$ are in
reasonable agreement, table~\ref{tab:energies} shows that the
corrections to the individual eigenvalues differ significantly.
In particular, DFT+$U$ consistently yields energies which are
$\sim$0.6~eV higher than in $GWU$.

In order to address the second point we consider the rutile structure. 
In our DFT and $GWU$ calculations the fundamental gap is direct 
at the $\Gamma$ point, therefore in table~\ref{tab:energies} we only
report our results for the $\Gamma_{24v}$ and $\Gamma_{25c}$ states.
We note, however, that the energies of the conduction band minima of rutile 
at $\Gamma$, $M$, and $R$ are within 60~meV from each other, and that
the location of the minimum is dependent upon the calculation method.
Given the small energy
difference between direct and indirect gaps in rutile, 
in order to establish the precise nature of the gap it would be necessary
to investigate the magnitude of the phonon-induced 
renormalization \cite{Marini2008,Giustino2010,Gonze2011,Marini2011}.
Table~\ref{tab:energies} shows that DFT+$U$ and $GWU$ calculations on rutile 
using $U_{\rm opt}$ lead to very similar band gaps, 2.83~eV and 2.85~eV, respectively.
This suggests that the optimal Hubbard parameter $U_{\rm opt}$ determined 
for anatase is reasonably transferable to rutile TiO$_2$.
Similarly to the case of anatase TiO$_2$, in the rutile polymorph
the corrections to the individual eigenvalues
calculated in DFT+$U$ are $\sim$0.5~eV higher than in $GWU$.
  \begin{table}
  \caption{
  Energies of critical points of anatase and rutile TiO$_2$ calculated using
  various methods: plain density-functional calculations (DFT),
  calculations including Hubbard-like corrections with $U$=7.5~eV
  (DFT+$U$), and $G_0W_0$ calculations starting from the DFT+$U$ eigenstates 
  and eigenvalues ($GWU$). For the sake of comparison with 
  other studies \cite{Chiodo2010, Kang2010, Mowbray2009, Thulin2008} we also report our results
  for $G_0W_0$ calculations starting from DFT eigenstates 
  and eigenvalues ($GW$). All values are in eV.
  The atomic positions and lattice parameters are taken from experiment \cite{Abrahams1971}.
  In the case of anatase we also give 
  the values calculated using our optimized lattice parameters
  between brackets. The choice of theoretical vs. experimental
  parameters lead to differences smaller than 0.1~eV.
  \vspace{0.3cm} \label{tab:energies}}
  \begin{tabular}{l r r r r r r r r}
  \br
  \multicolumn{9}{l}{Anatase TiO$_2$} \\
                    &  \multicolumn{2}{l}{DFT}  
                    &  \multicolumn{2}{l}{$GW$}
                    &  \multicolumn{2}{l}{DFT+$U$} 
                    &  \multicolumn{2}{l}{$GWU$} \\
  \mr
  $X_{24v}$         &  0.00 &(0.00) & -0.31 &(-0.30) &  0.48 &(0.49) & -0.14 &(-0.13) \\
  $\Gamma_{25c}$    &  2.08 &(2.06) &  3.39 & (3.38) &  3.75 &(3.70) &  3.13 & (3.07) \\
  $E_{\rm g}$       &  2.08 &(2.06) &  3.70 & (3.68) &  3.27 &(3.21) &  3.27 & (3.20) \\
  \br
  \vspace{-0.2cm}\\
  \multicolumn{9}{l}{Rutile TiO$_2$} \\
                    &  \multicolumn{2}{l}{DFT}  
                    &  \multicolumn{2}{l}{$GW$}
                    &  \multicolumn{2}{l}{DFT+$U$} 
                    &  \multicolumn{2}{l}{$GWU$} \\
  \hline
  $\Gamma_{24v}$    &  0.00 &       & -0.19 &        &  0.59 &       &  0.08 &        \\
  $\Gamma_{25c}$    &  1.82 &       &  3.21 &        &  3.42 &       &  2.93 &        \\
  $E_{\rm g}$       &  1.82 &       &  3.40 &        &  2.83 &       &  2.85 &        \\
  \br
  \end{tabular}
  \end{table}

In the case of anatase TiO$_2$ we are not aware of 
published inverse photoemission data on single crystals of anatase,
therefore it is not possible yet to perform a quantitative comparison between
our calculated quasiparticle energies and experiment. 
We note however that the measured optical gap of anatase 
(3.2~eV \cite{Kavan1996}) is compatible both with our $GW$ 
band gap (3.75~eV) and our $GWU$ band gap (3.27~eV).
In the case of rutile TiO$_2$ the quasiparticle band gap has 
been measured in the range $3.3\pm0.5$~eV \cite{Tezuka1994}.
This indicates that our $GW$ result (3.40~eV) and our $GWU$ calculation
starting from DFT+$U$ (2.85~eV) are both compatible with experiment.
The measured optical gap of rutile (3.0~eV \cite{Kavan1996}) 
is slightly larger than our calculated quasiparticle gap.

A key result of the present work is that, in the case of anatase and rutile
TiO$_2$, the use of DFT+$U$ as a starting Hamiltonian for calculating $GW$ 
quasiparticle energies leads to band gaps which are 0.4~eV smaller than
standard $GW$ calculations starting from DFT (\cite{Chiodo2010,Kang2010}
and table~\ref{tab:energies}).
This relatively large difference between $GW$ and $GWU$ raises the question
of which Hamiltonian is a 
better starting point for many-body perturbation theory
in strongly correlated systems.

Before concluding we note that, since in the approach described 
here the optimal Hubbard parameter 
is determined through an iterative procedure, this scheme
carries some similarities with self-consistent $GW$ methods \cite{Schilfgaarde2006}.
In the present work the self-consistent potential is constrained 
to be the DFT exchange and correlation potential plus an additional Hubbard-like term.
It would be interesting to apply the same concept to the case of
$GW$ calculations starting from hybrid functional 
calculations \cite{Perdew1996,Perdew1996b}, where instead of the Hubbard
$U$ we would optimize the fraction $\alpha$ of Hartree-Fock exchange.

In conclusion, we investigate the use of DFT+$U$ as a starting point for performing
many-body perturbation theory calculations on TiO$_2$.
We determine the Hubbard parameter of the non-interacting Hamiltonian
in such a way that a $GW$ calculation starting from the DFT+$U$ band structure 
leaves the fundamental band gap unchanged.
We find that the optimal Hubbard parameter of 7.5~eV yields the same
band gap of 3.27~eV in DFT+$U$ and $GWU$.
This value is considerably smaller than the value of 3.70~eV obtained
when applying $GW$ corrections to the DFT band structure.
The Hubbard parameter determined for anatase
is transferable to the rutile polymorph of TiO$_2$, 
and also in this case the calculated band gap is smaller than in standard
$GW$ calculations starting from DFT.
The difference between $GW$ and $GWU$ band gaps reported here calls for
detailed photoemission experiments to establish whether DFT+$U$ represents
a better starting point with respect to DFT for computing the $GW$ quasiparticle
energies of TiO$_2$.

\ack
We acknowledge support from the Engineering and Physical Sciences 
Research Council and from
the European Research Council under the European Community's Seventh Framework 
Programme (FP7/2007-2013)/ERC grant agreement no. 239578.

\end{document}